\documentclass[preprintnumbers,amsmath,amssymb,showpacs,twocolumn]{revtex4}

\usepackage{graphicx}
\usepackage{dcolumn}
\usepackage{bm}

\begin{document}

\title{ Quantum secret sharing between \emph{m}-party and \emph{n}-party  with six states}

\author{Ting Gao$^{1,2}$, Feng-Li Yan$^{2,3}$, You-Cheng Li$^{2,3}$}

\affiliation {$^1$ College of Mathematics and Information Science, Hebei Normal University, Shijiazhuang 050016, China\\
$^2$ CCAST (World Laboratory), P.O. Box 8730, Beijing 100080, China\\
$^3$ College of Physics  and Information Engineering, Hebei Normal
University, Shijiazhuang 050016, China}

\date{\today}

\begin{abstract}
We propose a quantum secret sharing scheme between $m$-party  and $n$-party  using three conjugate bases, i.e.
six states. A sequence of single photons, each of which is prepared in one of the six states, is used directly
to encode classical information in the quantum secret sharing process. In this scheme, each of all $m$ members
in group 1 choose randomly their own secret key individually and independently, and then directly encode their
respective secret information on the states of single photons via unitary operations, then the last one (the
$m$th member of group 1) sends $1/n$ of the resulting qubits to each of group 2. By measuring their respective
qubits, all members in group 2 share the secret information shared by all members in group 1. The secret message
shared by group 1 and group 2 in such a way that neither subset of each group nor the union of a subset of group
1 and a subset of group 2 can extract
  the secret message, but each whole group (all the members of each group) can. The scheme is asymptotically
   100$\%$ in efficiency. It  makes the Trojan horse
attack with a multi-photon signal, the fake-signal attack with EPR pairs, the attack with single photons, and
the attack with invisible photons to be nullification. We  show that it is   secure and has an advantage over
the one based on two conjugate bases.  We also give  the upper bounds of the average success probabilities for
dishonest agent eavesdropping encryption using the fake-signal attack with any two-particle entangled states.
This protocol is feasible with present-day technique.
\end{abstract}


\maketitle

\section{Introduction}

 Secret sharing is a powerful technique in computer science, which enables secure and
robust communication in information networks, such as the internet,
telecommunication systems and distributed computers. The security of
these networks can be enhanced using quantum resources to protect
the information. Such schemes have been termed {\it quantum secret
sharing}. There are many kinds one-to-multi-party  quantum secret
sharing schemes, such as with entanglement \cite{HBB, TZG, KKI,
Gottesman, NQI, CGL, KBB, BK, XLDP, BD} and without entanglement
\cite{GG03}. The basic idea of secret sharing in the simplest case
is that a secret of one person, say Alice, is shared between other
two persons, say Bob and Charlie, in such a way  that it can only be
reconstructed if both collaborate. In a more general case, notably
for secure key management, a $t$-{\it out}-{\it of}-$n$ protocol (or
$(t, n)$-threshold scheme) with $1\leq t\leq n$ spreads a secret to
$n$ participants in a way that any $t$ participants can reconstruct
it \cite{TOI}. Lance {\it et al.} have reported an experimental
demonstration of a (2,3) threshold quantum secret sharing scheme
\cite {LSBSL}. The interesting aspect of exploring quantum mechanics
for secret sharing is that it allows for unconditionally secure
distribution of the information to the participants.

In  Ref.\cite{HBB}, Hillery {\it et al.} proposed one-to-two-party and one-to-three-party  secret sharing
schemes via maximally entangled three-particle and four-particle states, respectively. Xiao {\it et al.}
\cite{XLDP} reformulated the protocol \cite {HBB} in a simple mathematical terms and generalized the protocol
\cite {HBB} into arbitrary number parties case.
 The main purpose of Ref.\cite{KKI}
is to show that it is also possible to realize secret sharing as in Ref.\cite{HBB} with two-particle quantum
entanglement. At the same time, Karlsson {\it et al.} \cite{KKI} also presented a detailed discussion of how to
detect eavesdropping, or how to detect a dishonest party in the protocols.
 Quantum secret sharing protocol provides for secure secret sharing by enabling one to
determine whether an eavesdropper has been active during the secret sharing procedure. But it is not easy to
implement such multi-party secret sharing tasks \cite {HBB, KKI, XLDP}, since the efficiency of preparing even
tripartite or four-partite entangled states is very low \cite {BPDWZ, PDGWZ}, at the same time  the efficiency
of the existing quantum secret sharing protocols using quantum entanglement can only approach $50\%$. Recently,
a scheme for quantum secret sharing without entanglement has been proposed by Guo and Guo \cite {GG03}. They
presented an idea to directly encode the qubit of quantum key distribution and accomplish one splitting a
message into many parts to achieve one-to-multi-party secret sharing only by product states. The theoretical
efficiency is doubled to approach $100\%$.

In the modern society, signatures on documents, authentications, encryptions, and decryptions are often needed
by more than one person, especially by all persons of two groups. Therefore, the secret sharing between many
parties and many parties is indeed required. The advantage of secret sharing between many parties and many
parties is that honest agents can keep the dishonest ones from doing any damage when they  appear in the process
for business. More recently, we suggest a quantum secret sharing scheme employing two conjugate bases, i.e. four
states, of single qubits to achieve the secret sharing between multi-party  and multi-party with a sequence of
single photons \cite{YGpra1}.
 Deng {\it et al.} \cite {DYLLZG} and Li {\it et
al.} \cite {LiChangHwang} showed the weakness of our protocol
\cite{YGpra1}. We proposed an improved protocol \cite{YGL}, which
can avoid the  flaws and is secure against the attack with invisible
photons \cite{caiqingyu} and the fake-signal attack with any
two-particle entangled states (the special case of which is the
fake-signal attack with EPR pairs \cite{DLZZ2}).

In this paper, we propose a  quantum secret sharing scheme between $m$-party and $n$-party  by using three
rather  than two conjugate bases. In this scheme, all \emph{m} parties in group 1 select their own secrets
individually and independently, and split their respective secret information among the whole group 2 ($n$
parties) in such a way that neither part members of either group 1 or group 2 nor the union of a subset of group
1 and a subset of group 2 has any knowledge of the combination of all senders (group 1), but only  by working
together can all members of each group jointly determine what the combination of all senders (group 1) is. In
this case it is the secret information of each person in group 1 that has been split up into $n$ pieces, no one
of which separately contains the original information, but whose combination does. We show  that this
\emph{m}-to-\emph{n}-party scheme  is more secure than the one  in \cite{YGpra1, YGL} based on two conjugate
bases, i.e. four states, such as its security against  the Trojan horse attack \cite {DYLLZG}, the two attacks
stated in Ref. \cite {LiChangHwang}, the attack with invisible photons \cite {caiqingyu}, and the attack with
EPR pairs \cite {DLZZ2}. In fact, we show that this protocol is secure against more general attack (the attack
with any two-particle entangled state) than the attack with EPR pairs.  The advantage of the present scheme is
that two honest agents, one in group 1 and the other in group 2, can keep the dishonest ones ( other $m+n-2$
members) and the $m+n+1$ party (an 'external' Eavesdropper, Eve) from doing any damage. That is, this protocol
is secure as long as there is one agent at each group being honest. Comparing with the efficiency $50\%$
limiting for the existing quantum secret sharing protocols with quantum entanglement, the present scheme can
also be $100\%$ efficient in principle.

\section{quantum key sharing between multi-party and multi-party based on six states}

Let Alice 1, Alice 2, $\cdots$, Alice $m$, and Bob 1, Bob 2,
$\cdots$, Bob $n$ be  respective all members of group 1 and group 2.
$m$ parties of group 1 want quantum key sharing with $n$ parties of
group 2 such that neither  part of each group  nor the union of a
part of group 1 and a part of group 2 knows the key, but only all
members of each group can collaborate to determine what the string
(key) is. Next we put forward an $m$-to-$n$-party quantum secret
sharing scheme  by using of three conjugate bases, or six states, to
achieve the aim mentioned above---the secret sharing between $m$
parties and $n$ parties. Let us see how this works in detail.

M1. Alice 1 creates a random $nN$ bit  string $A_1$ and a random
$nN$ trit string $B_1$, where $a^1_k$ and $b^1_k$ are uniformly
chosen from $\{0, 1\}$ and $\{0, 1, 2\}$, respectively. She then
encodes these strings as a block of $nN$ qubits ($nN$ single
photons),
\begin{eqnarray}\label{Alice1}
   |\Psi^1\rangle = \otimes_{k=1}^{nN}|\psi_{a^1_kb^1_k}\rangle,
   \end{eqnarray}
where $a^1_k$ is the $k$th bit of $A_1$ (and similar for $B_1$).
Each qubit $|\psi_{a^1_kb^1_k}\rangle$ is in one of the six states
\begin{equation}\label{6states}
\begin{array}{ll}
 |\psi_{00}\rangle=|0\rangle, &  |\psi_{10}\rangle =|1\rangle, \\
 |\psi_{01}\rangle=|+\rangle=\frac{|0\rangle+|1\rangle}{\sqrt{2}}, & |\psi_{11}\rangle=|-\rangle=\frac{|0\rangle-|1\rangle}{\sqrt{2}}, \\
 |\psi_{02}\rangle=|+y\rangle=\frac{|0\rangle+\texttt{i}|1\rangle}{\sqrt{2}}, & |\psi_{12}\rangle=|-y\rangle=\frac{|0\rangle-\texttt{i}|1\rangle}{\sqrt{2}}.
\end{array}
\end{equation}
The value of $b^1_k$ determines the basis. If $b^1_k$ is 0 then
$a^1_k$ is encoded in the $Z$ basis $\{|0\rangle, |1\rangle\}$; if
$b^1_k$ is 1 then $a^1_k$ is encoded in the $X$ basis
$\{|+\rangle, |-\rangle\}$; if $b^1_k$ is 2 then $a^1_k$ is
encoded in the $Y$ basis $\{|+y\rangle, |-y\rangle\}$.  Note that
the six states are not all mutually orthogonal, therefore no
measurement can distinguish between  all of them with certainty.
 Alice 1 then sends the sequence of $nN$ single photons
 to Alice 2 over their public quantum communication channel.

M2. When Alice 2 receives the  $nN$ qubits, she chooses randomly a
large subset of photons as the samples for eavesdropping check.
First, she uses a special filter to prevent the invisible photons
from entering the operation system,  splits each sample signal with
a photon number splitter (PNS: 50/50), and then measures each signal
in the measurement basis (MB) $Z$, or $X$, or $Y$ at random.
Obviously if two or more photons in one signal are detected, then
Alice 2 aborts the communication. Moreover, she
 analyzes the error rate $\varepsilon_s$ of the samples by
 requiring Alice 1 to tell her the original states of
the samples. If the error rate is higher than the threshold chosen by all Alices and Bobs \cite{NC}, Alice 2
aborts the communication, otherwise she goes ahead. After Alice 2's checking, the number of the remaining
unchecked qubits (photons) must be less than $nN$. However, for convenience, we suppose that Alice 2 still has
the $nN$ photons.

  Alice 2 selects two random $nN$-trit strings $A_2$ and $B_2$.
She
  performs the
operation $\sigma_0=I=|0\rangle\langle0|+|1\rangle\langle 1|$,
$\sigma_1=\texttt{i}\sigma_y=|0\rangle\langle1|-|1\rangle\langle0|$ or
$\sigma_2=\sigma_z=|0\rangle\langle0|-|1\rangle\langle 1|$  on each qubit $|\psi_{a^{1}_kb^{1}_k}\rangle$  if
the corresponding trit value $a^2_k$ of $A_2$ is $0$, $1$ or $2$, respectively. Then, she applies a unitary
operator $U_0=I$,
$U_1=\frac{1}{\sqrt{2}}(|0\rangle+|1\rangle)\langle0|-\frac{\texttt{i}}{\sqrt{2}}(|0\rangle-|1\rangle)\langle1|$,
or
$U_2=\frac{1}{\sqrt{2}}(|0\rangle+\texttt{i}|1\rangle)\langle0|+\frac{1}{\sqrt{2}}(|0\rangle-\texttt{i}|1\rangle)\langle1|$
on the $k$th photon depending on $b^2_k=0$, $1$, or $2$, respectively. We denote each of the resulting qubit
states as $|\psi_{a^2_kb^2_k}\rangle$. After that, Alice 2 inserts randomly $N_2$ decoy single photons  into
$nN$ photons encoded by her, where each of  the decoy single photons is randomly in one of the states in
Eq.({\ref{6states}}). Then she sends Alice 3 the resulting $(nN+N_2)$-qubit state
$|\Psi^2\rangle=\otimes_{k=1}^{nN+N_2}|\psi_{a^2_kb^2_k}\rangle$. Notice that $|\psi_{a^2_kb^2_k}\rangle$ is
determined by ${a_k^1, b_k^1, a_k^2, b_k^2}$.

The nice feature of  the unitary transformations
$\texttt{i}\sigma_y, \sigma_x, \sigma_z$ is that they leave bases
$Z$, $X$ and $Y$ unchanged,  and  each of them flips the states in
two measurement bases. For example,  $\sigma_1=i\sigma_y$ flips the
states in both bases $X$ and $Z$ such that
\begin{equation}\label{Y}
    \begin{array}{ll}
      \sigma_1|0\rangle=-|1\rangle, & \sigma_1|1\rangle=|0\rangle, \\
      \sigma_1|+\rangle=|-\rangle, & \sigma_1|-\rangle=-|+\rangle.
     \end{array}
\end{equation}
The operators $U_1$ and $U_2$ cyclically permute the three bases
$Z$, $X$ and $Y$ such that
\begin{equation}\label{U1}
U_1: \{|0\rangle, |1\rangle\}\rightarrow \{|+\rangle, |-\rangle\}
\rightarrow\{|+y\rangle, |-y\rangle\}\rightarrow\{|0\rangle,
|1\rangle\}
\end{equation}
 and
\begin{equation}\label{U2}
U_2: \{|0\rangle, |1\rangle\}\rightarrow\{|+y\rangle,
|-y\rangle\}\rightarrow \{|+\rangle,
|-\rangle\}\rightarrow\{|0\rangle, |1\rangle\}.
\end{equation}

M3. Alice $i$ ($3\leq i\leq m$) operates the qubits like Alice 2 does. That is, first, she select a large subset
of photons at random as the samples for eavesdropping check. For determining the error rate of the samples, all
the members before Alice $i$ must tell Alice $i$ the original state or the operations they chose in a random
sequential order. Because of nice feature of the unitary transformations $\texttt{i}\sigma_y$, $\sigma_x$,
$\sigma_z$, $U_1$, $U_2$, each qubit $|\psi_{a^i_kb^i_k}\rangle$ of the resulting $nN+N_{i-1}$ qubit product
state $\otimes_{k=1}^{nN+N_{i-1}}|\psi_{a^i_kb^i_k}\rangle$ is the eigenstate of $\sigma_z$, $\sigma_x$, or
$\sigma_y$ if
 $(b_k^1+b_k^2+\cdots+b_k^i)\text{mod}3=0$, 1, or 2,
respectively.  Evidently, $|\psi_{a^i_kb^i_k}\rangle$ is govern by
$a_k^1,a_k^2,\cdots, a_k^i,b_k^1,b_k^2,\cdots,b_k^i$. Here
$i=3,4,\cdots, m$. Second, Alice $i$ generates two random
$(nN+N_{i-1})$-trit strings $A_i$ and $B_i$,    applies
$\sigma_0$, $\sigma_1$, or $\sigma_2$ on the $k$-th photon
depending $a_k^i=0$, 1, or 2 and performs $U_0$, $U_1$, or $U_2$
on the resulting state of the $k$-th photon according to
$b_k^i=0$, 1, 2. Finally, Alice $i$ inserts randomly $N_i-N_{i-1}$
decoy single photons into $nN+N_{i-1}$ photons encoded by her.

M4. Alice $i$ ($3\leq i\leq m-1$) sends the resulting $nN+N_i$ qubit product state
$|\Psi^i\rangle=\otimes_{k=1}^{nN+N_i}|\psi_{a^i_kb^i_k}\rangle$ to Alice $i+1$.  Without loss of generality,
we can suppose that $nN+N_m=n\overline{N}$ (Alice $m$ can manage it). Alice $m$ sends $\overline{N}$-qubit
product states $|\Psi_1^m\rangle=\otimes_{j=0}^{\overline{N}-1}|\psi_{a^m_{nj+1}b^m_{nj+1}}\rangle$,
$|\Psi_2^m\rangle=\otimes_{j=0}^{\overline{N}-1}|\psi_{a^m_{nj+2}b^m_{nj+2}}\rangle$, $\cdots$,
 $|\Psi_n^m\rangle=\otimes_{j=0}^{\overline{N}-1}|\psi_{a^m_{nj+n}b^m_{nj+n}}\rangle$ of the resulting
           $n\overline{N}$-qubit state
$|\Psi^m\rangle=\otimes_{k=1}^{n\overline{N}}|\psi_{a^m_kb^m_k}\rangle$ to Bob 1, Bob 2, $\cdots$, Bob $n$, respectively.

M5. When all Bob 1, Bob 2, $\cdots$, and Bob $n$ have  received their respective $n\overline{N}$ qubits, each of
them   first randomly and independently chooses enough  photons as samples and measures each of them in MB $Z$,
or $X$ or $Y$ at random. Then they ask Alice 1, Alice 2, $\cdots$, and Alice $m$ to announce publicly the
$a^i_{k_t}$ and $b^i_{k_t}$ of the samples in a random sequential order. Here $k_t$ is the label of the sample
chosen for eavesdropping check, and $i=1,2,\cdots, m$. After that Bobs publish their measurement outcomes and
the measurement bases. All Alices and Bobs discard all check photons except those for which Bobs measured in the
MB  $Z$, $X$  or $Y$  according to $(\sum_{i=1}^mb^i_{k_t})\text{mod}3=0$, 1 or 2, and compare the values of
their remaining check photons. If the error rate of the remaining samples  is reasonably less than a threshold,
then they continue to the quantum communication. Otherwise they abort it.

M6. Alice 1, Alice 2, $\cdots$, and Alice $m$ ask all Bobs to discard the decoy photons that are not chosen for
eavesdropping check, and then publicly announce the strings $B_1$, $B_2$, $\cdots$, and $B_m$ at random,
respectively. Bob 1, Bob 2, $\cdots$, and Bob $n$ then measure each qubit of their respective strings in MB $Z$
or $X$ or $Y$ according to the result of addition modulo 3 of corresponding trit values of strings $B_1$, $B_2$,
$\cdots$, $B_m$. Thus,  if $(\sum_{i=1}^m b^i_{nj+l})\text{mod}3=0$, then Bob $l$ measures
$|\psi_{a^m_{nj+l}b^m_{nj+l}}\rangle$ in the $Z$ basis; if $(\sum_{i=1}^m b^i_{nj+l})\text{mod}3=1$, he measures
in the $X$ basis; if $(\sum_{i=1}^m b^i_{nj+l})\text{mod}3=2$, he measures in the $Y$ basis. After his
measurement, Bob $l$ can extract out the combination of all Alices's encoding information.

Let measurement result of Bob $l$ be $C_l=\{c_{nj+l}\}_{j=0}^{\overline{N}-1}$ and the combination of all
Alices's  encoding information be $A=\{a_k\}_{k=1}^{n\overline{N}}$, where $a_k$ is determined by all $a_k^i$
and $b_k^i$, and $a_k$ and $c_{nj+l}$ are 0 or 1, corresponding to the +1 and -1 eigenstates of $\sigma_z$,
$\sigma_x$, and $\sigma_y$.
 Clearly, if there
are no eavesdropper and noise, there must be $c_{nj+l}=a_{nj+l}$.
Here $l=1, 2, \cdots, n$ and $i=1, 2, \cdots, m$.

M7. All Alices and Bobs perform some tests to determine how much noise or eavesdropping happened during their
communication. Alice 1, Alice 2, $\cdots$, and Alice $m$ select some photons $nj_r+l$ (of their $n\overline{N}$
photons) at random, and publicly announce the selection.
 All Bobs and all Alices then publish and compare
the values of these checked bits. If they find too few $a_{nj_r+l}=
c_{nj_r+l}$, then they abort and re-try the protocol from the start.

M8. The XOR results $\oplus^n_{l=1}c_{nj_s+l}$  of Bob $l$'s corresponding bits $c_{nj_s+l}$ of the rest
unchecked  photons $nj_s+l$ of $\otimes_{j=0}^{\overline{N}-1}|\psi_{a^m_{nj+1}b^m_{nj+1}}\rangle$,
$\otimes_{j=0}^{\overline{N}-1}|\psi_{a^m_{nj+2}b^m_{nj+2}}\rangle$, $\cdots$,
 $\otimes_{j=0}^{\overline{N}-1}|\psi_{a^m_{nj+n}b^m_{nj+n}}\rangle$ can be used as raw keys
for secret sharing between all Alices and all Bobs.

Remark 1. It is necessary for Alice 2, Alice 3, $\cdots$, Alice $m$ make a eavesdropping check before they
operates the  photon signals, otherwise dishonest agent Alice $i_0$ can obtain secret messages of Alice $i_0+1$,
$\cdots$, Alice $m$ with a Trojan horse attack and invisible photons attack. The reasons are similar to that
stated in Ref. \cite {DLZZ, DYLLZG}.

Remark 2. By random sampling in M5, the security flaw indicated in
\cite{LiChangHwang} can be avoided in the present protocol. That is,
for this protocol, the two types attacks, the attack with EPR pairs
and the attack with single photons, proposed in \cite{LiChangHwang}
are of no effect. The reason is as follows.
 Evidently, Bobs'  measurements collapse
   the check samples into the states of a single particle.
That is, all Bobs' measurements remove the entanglements between check photons and other eavesdropping
particles, which correspond to that the attacker Alice $i_0$ sends Bobs single quantum states Eq.(\ref{6states})
whether in the attack with EPR pairs or the attack with single photons. Apparently, we can find out the attacker
in the attack with single photons via Alices' publishing  their respective encoding information in a random
sequential order.
 Since Bob $l$ asks Alices to announce the $a^i_{k_t}$
and $b^i_{k_t}$ in a random sequential order, the attacker will not be the last one to answer the Bob $l$'s
enquiry with a probability $\frac {m-1}{m}$. If there is a Alice to be asked after the attacker Alice $i_0$,
Alice $i_0$ can not distinguish the quantum state intercepted by her with certainty,  and can only guess
$a^{i_0}_{k_t}$ and $b^{i_0}_{k_t}$ to answer the inquiry.  It is not difficult to deduce that the error rate of
the samples that Bobs measured in MB $Z$, $X$, or $Y$ corresponding to $(\sum_{i=1}^m b_{k_t}^i){\rm mod}3=0,
1$, or 2 is more than $\frac {m-1}{2m}$.

Remark 3. It is very nice that this  scheme can make the fake-signal
attack with any two-particle entangled state (the special case of
which is the fake-signal attack with EPR pairs \cite{DLZZ2}) to be
nullification. The argument goes as follows:

Suppose that the eavesdropper Alice $i_0$  (who could  be any dishonest one of Alices) generates $nN+N_{i_0}$
general EPR pairs in the state $\otimes_{k=1}^{nN+N_{i_0}}|\Psi_k\rangle$. Here
\begin{equation}
 |\Psi_k\rangle=|\Psi\rangle=|0\rangle_A|\alpha\rangle_E+|1\rangle_A|\beta\rangle_E,
 \end{equation}
$|\alpha\rangle_E$ and $|\beta\rangle_E$  are unnormalized
  states of the $S$-level ($S\geq 2$) particle $E$. Note that   the special case
  of $|\Psi\rangle$ is an EPR pair (when $\langle\alpha|\beta\rangle = \langle\beta|\alpha\rangle=0,
  \langle\alpha|\alpha\rangle =\langle\beta|\beta\rangle=\frac{1}{2}$,
$|\Psi_k\rangle=|\Psi\rangle$ is an EPR pair). Alice $i_0$ keeps the second particle $E$ of each
$|\Psi_k\rangle$,
 replaces the original
single photons in the state $|\Psi^{i_0}\rangle$ with the first particle $A$ of each $|\Psi_k\rangle$ and
 sends the sequence $S_A$ of $nN+N_{i_0}$ qubit A to Alice $i_0+1$ (this is a more general
  situation than that in \cite{DLZZ2}).

 If $|\Psi\rangle$ is not a two-particle maximally
entangled state (EPR pair), then Alice $i_0$ can not but makes mistakes  in M2 (if $i_0=1$) or M3 (if $i_0>1$),
because Alice $i_0$ can not distinguish between $|\alpha\rangle$ and $|\beta\rangle$, between
$|\alpha\rangle+|\beta\rangle$ and $|\alpha\rangle-|\beta\rangle$, and between
$|\alpha\rangle+\texttt{i}|\beta\rangle$ and $|\alpha\rangle-\texttt{i}|\beta\rangle$ perfectly. Thus  Alice
$i_0+1$ can detect the cheating of Alice $i_0$  in M2 (if $i_0=1$) or M3 (if $i_0>1$). Next we only assume that
$|\Psi\rangle$ is an EPR pair.  Alice $i_0+1$ in the step M2 (if $i_0=1$) or M3 (if $i_0>1$) cannot detect this
cheating as Alice $i_0$ is able to produce no errors in the results if Alice $i_0$ is asked to announce her
encryption $a_s^{i_0}$ and $b_s^{i_0}$ of the samples after Alice 1, $\cdots$, Alice $i_0-1$. But if Alice $i_0$
is not the last to announce her encoding information, then her cheating introduces errors and can be found out
by Alice $i_0+1$ in M3 (if $i_0>1$) without fail. However, when the dishonest Alice $i_0$ is Alice 1,  this
cheating of her cannot be found out by Alice 2 as it does not introduce errors in the results.

 Alice 1 intercepts $S_A$ while it was sent to Alice $i_1$
($2<i_1\leq m$) or Bobs (if Alice 1 never intercepts $S_A$,  then she can not obtain any information, although
this kind of eavesdropping can not be found in the eavesdropping check. So it does not make any sense for Alice
1 to do this kind of eavesdropping ).

The object of Eve is to obtain all Alices' encoding information
$A_i$ and $B_i$. In order to achieve this purpose, Alice 1 must
manage to distinguish nine unitary operations $\sigma_0$,
$\sigma_1$, $\sigma_2$, $U_1$, $U_1\sigma_1$, $U_1\sigma_2$, $U_2$,
$U_2\sigma_1$, and $U_2\sigma_2$ (if $i_1=3$) or more than these
nine operations (if $i_1\geq 4$). That is, Alice 1 must manage to
distinguish the following 9 states
 \begin{eqnarray}\label{9states}
 &&|\chi_1\rangle =|\Psi\rangle=|0\rangle|\alpha\rangle+|1\rangle|\beta\rangle,\nonumber\\
 &&|\chi_2\rangle =\sigma_1|\Psi\rangle=-|1\rangle|\alpha\rangle+|0\rangle|\beta\rangle,\nonumber\\
&&|\chi_3\rangle =\sigma_2|\Psi\rangle=|0\rangle|\alpha\rangle-|1\rangle|\beta\rangle,\nonumber\\
 &&|\chi_4\rangle =U_1|\Psi\rangle\nonumber
 = \frac{1}{\sqrt 2}(|0\rangle+|1\rangle)|\alpha\rangle-\frac{\texttt{i}}{\sqrt 2}(|0\rangle
- |1\rangle)|\beta\rangle,\nonumber\\
 &&|\chi_5\rangle=U_1\sigma_1|\Psi\rangle =\frac{\texttt{i}}{\sqrt 2}(|0\rangle-|1\rangle)|\alpha\rangle+\frac{1}{\sqrt 2}(|0\rangle+|1\rangle)|\beta\rangle,\nonumber\\
  &&|\chi_6\rangle=U_1\sigma_2|\Psi\rangle =\frac{1}{\sqrt 2}(|0\rangle+|1\rangle)|\alpha\rangle+\frac{\texttt{i}}{\sqrt 2}(|0\rangle-|1\rangle)|\beta\rangle,\nonumber\\
 &&|\chi_7\rangle=U_2|\Psi\rangle=\frac{1}{\sqrt 2}(|0\rangle+\texttt{i}|1\rangle)|\alpha\rangle
 +\frac{1}{\sqrt 2}(|0\rangle-\texttt{i}|1\rangle)|\beta\rangle,\nonumber\\
  &&|\chi_8\rangle=U_2\sigma_1|\Psi\rangle =-\frac{1}{\sqrt 2}(|0\rangle-\texttt{i}|1\rangle)|\alpha\rangle+\frac{1}{\sqrt
 2}(|0\rangle+\texttt{i}|1\rangle)|\beta\rangle,\nonumber\\
 &&|\chi_9\rangle=U_2\sigma_2|\Psi\rangle =\frac{1}{\sqrt 2}(|0\rangle+\texttt{i}|1\rangle)|\alpha\rangle-\frac{1}{\sqrt
 2}(|0\rangle-\texttt{i}|1\rangle)|\beta\rangle,\nonumber\\
 \end{eqnarray}
 or even more, where $\sigma_i$ and $U_i\sigma_i$ act on the first particle, and $|\Psi\rangle$ is an EPR pair.
These states belongs to the four dimension subspace $W$ of the entire Hilbert space $\mathcal{H}$ spanned by
$|0\rangle|\alpha\rangle$, $|0\rangle|\beta\rangle$, $|1\rangle|\alpha\rangle$ and $|1\rangle|\beta\rangle$, and
it is impossible for more than four states being mutually orthogonal, therefore these  states can not be
reliably distinguished \cite{NC}. Thus Alice 1's eavesdropping can be found in M3 by Alice $i_1$ or in M5 by all
Bobs. Therefore the present quantum secret sharing protocol is secure against the fake-signal attack with any
two-particle entangled state (the special case of which is EPR pairs).

On the other hand, in the present protocol, Alices  insert randomly decoy single photons into the signal photons in M2 or M3. The eavesdropping
check on the decoy single photons is the same as that on the signal photons. That is, first Alice $i$ measures (or all Bobs measure) each decoy
single photon in MB $Z$, $X$ or $Y$ at random, then she asks (they ask) Alice 1, Alice 2, $\cdots$, Alice $i-1$ (all Alices) to tell her (them)
their encoding information $a^t_s, b^t_s$ of the samples in a random sequential order.  Note that there is at least one honest agent  in one
communication group. Therefore the dishonest agent can be found by the eavesdropping checks on the decoy photons by the honest agents. The
principle of the checking procedures is the same as that in six states quantum key distribution protocol \cite{BBBW, Bru}.

 Next we give the upper bounds of the average success probabilities  of  two cases.

 Case I.  The upper bound of the average success probability
distinguishing the nine states in Eq.(\ref{9states}), where $|\Psi\rangle$ is a general two-particle entangled
state. Since  the non-orthogonal states can not be reliably distinguished \cite{NC}, distinguishing the nine
states is equivalent to find $|\alpha\rangle$ and $|\beta\rangle$ such that
\begin{equation}\label{orthogonality}
\langle\chi_i|\chi_j\rangle=\delta_{ij}.
\end{equation}
However, there are no $|\alpha\rangle, |\beta\rangle$ satisfying
Eq.(\ref{orthogonality}), as even there are no $|\alpha\rangle,
|\beta\rangle$ satisfying the following equations
\begin{equation}
\langle \chi_1|\chi_2\rangle=\langle \alpha|\beta\rangle-\langle
\beta|\alpha\rangle=0,
\end{equation}
\begin{equation}
\langle \chi_1|\chi_3\rangle=\langle \alpha|\alpha\rangle-\langle
\beta|\beta\rangle=0,
\end{equation}
\begin{equation}
\langle \chi_1|\chi_4\rangle=\frac {1}{\sqrt 2}(1+\texttt{i})\langle
\alpha|\alpha\rangle-\frac {1}{\sqrt 2}(1-\texttt{i})\langle
\alpha|\beta\rangle=0,
\end{equation}
\begin{equation}
\langle \chi_1|\chi_5\rangle=\frac {\texttt{i}}{\sqrt 2}\langle
\alpha|\alpha\rangle-\frac {1}{\sqrt 2}\langle
\alpha|\beta\rangle-\frac {\texttt{i}}{\sqrt 2}\langle
\beta|\alpha\rangle+\frac {1}{\sqrt 2}\langle \beta|\beta\rangle=0.
\end{equation}
Thus, Eve can not
 reliably distinguish the nine states in Eq.(\ref{9states}).

We calculate the upper bound of the maximal success probability for Eve unambiguously discriminating 9 states in
Eq.(\ref{9states}). Let
\begin{equation}
  \begin{array}{ccl}
  \frac {1}{\sqrt 2}\langle \alpha|\beta\rangle & = & x+\texttt{i}y, \\
  \frac {1}{\sqrt 2}\langle \alpha|\alpha\rangle & = & z, \\
  \frac {1}{\sqrt 2}\langle \beta|\beta\rangle & = & t,
  \end{array}
\end{equation}
then
\begin{eqnarray}
\frac {1}{\sqrt 2}\langle \beta|\alpha\rangle=x-\texttt{i}y,\nonumber\\
 z>0, ~~t>0, ~~z+t=\frac {1}{\sqrt 2},
\end{eqnarray}

\begin{eqnarray}
&& \frac {1}{2}\sum^9_{\substack{i,j=1 \\ i\neq j}}
|\langle\chi_i|\chi_j\rangle|\nonumber\\
=&& 6\sqrt {2}|y|+ 3\sqrt {2}|z-t|+6\sqrt {2}|x|\nonumber\\
&&+4\sqrt {(x+y+z)^2+(x+y-t)^2}\nonumber\\
&&+6\sqrt {(x-y-z)^2+(x-y+t)^2}\nonumber\\
&&+4\sqrt {(x-y+z)^2+(x-y-t)^2}\nonumber\\
&&+4\sqrt {(x+y-z)^2+(x+y+t)^2}\nonumber\\
&&+{\sqrt 2}\sqrt {\frac {1}{2}+(2x+2y+z-t)^2}\nonumber\\
&&+\frac {3}{\sqrt 2}\sqrt {\frac {1}{2}+(2x-2y-z+t)^2}\nonumber\\
&&+{\sqrt 2}\sqrt {\frac {1}{2}+(2x-2y+z-t)^2}\nonumber\\
&&+{\sqrt 2}\sqrt {\frac {1}{2}+(2x+2y-z+t)^2}.
\end{eqnarray}

By simply calculating, we obtain that
\begin{equation}
\sum_{\substack{i,j=1 \\
i\neq j}}^9|\langle\chi_i|\chi_j\rangle|\geq 27.
\end{equation}
Moreover, the equality holds if
\begin{equation}
    x=y=0, t=z=\frac{1}{2\sqrt{2}},
\end{equation}
That is, the minimum of $\sum_{\substack{i,j=1 \\
i\neq j}}^9|\langle\chi_i|\chi_j\rangle|$ occurs at
\begin{equation}
   \langle\alpha|\beta\rangle=0,
   \langle\alpha|\alpha\rangle=\langle\beta|\beta\rangle=\frac{1}{2}.
\end{equation}

The average success probability $P_1$ \cite{ZFSY} for unambiguous
identification of the nine states in Eq.(\ref{9states}) is
\begin{equation}\label{p1}
P_1\leq 1-{\frac {1}{9-1}}\sum^9_{\substack{i,j=1 \\
i\neq j}}\sqrt {
{\frac{1}{9}\times\frac{1}{9}}}|\langle\chi_i|\chi_j\rangle|=1-\frac
{3}{8}=\frac {5}{8}.
\end{equation}

Case II. The upper bound of the average success probability
classifying the following three sets
\begin{eqnarray}
 &&\{|\chi_{11}\rangle, |\chi_{12}\rangle, |\chi_{13}\rangle\},\nonumber\\
&&\{|\chi_{21}\rangle, |\chi_{22}\rangle, |\chi_{23}\rangle\},\\
&&\{|\chi_{31}\rangle, |\chi_{32}\rangle,
|\chi_{33}\rangle\}\nonumber
\end{eqnarray}
 for gaining secret information $A_i$, where
\begin{eqnarray}
&& |\chi_{11}\rangle=|\chi_1\rangle, |\chi_{12}\rangle=|\chi_4\rangle, |\chi_{13}\rangle=|\chi_7\rangle,\nonumber\\
&& |\chi_{21}\rangle=|\chi_2\rangle, |\chi_{22}\rangle=|\chi_5\rangle, |\chi_{23}\rangle=|\chi_8\rangle,\nonumber\\
&& |\chi_{31}\rangle=|\chi_3\rangle, |\chi_{32}\rangle=|\chi_6\rangle, |\chi_{33}\rangle=|\chi_9\rangle.
\end{eqnarray}

It is not difficult to derive
\begin{eqnarray}
&&\sum^3_{\substack{i,j=1 \\ i\neq j}}\sum_{k,l=1}^3\sqrt {\frac
{\eta_{ik}\eta_{jl}}{
{(N-m_i)(N-m_j)}}}|\langle\chi_{ik}|\chi_{jl}\rangle|\nonumber\\
=&&\sum^3_{\substack{i,j=1 \\ i\neq j}}\sum_{k,l=1}^3\sqrt {\frac
{\frac{1}{9}\times\frac{1}{9}}{
{(9-3)\times (9-3)}}}|\langle\chi_{ik}|\chi_{jl}\rangle|\nonumber\\
=&& \frac {1}{27}(6\sqrt {2}|y| +3\sqrt {2}|z-t|+6\sqrt {2}|x|\nonumber\\
&&+2\sqrt {(x+y+z)^2+(x+y-t)^2}\nonumber\\
&& +6\sqrt {(x-y-z)^2+(x-y+t)^2} \nonumber\\
&&+2\sqrt {(x-y+z)^2+(x-y-t)^2}\nonumber\\
&& +2\sqrt {(x+y-z)^2+(x+y+t)^2}\nonumber\\
&& +\frac {3}{\sqrt 2}\sqrt {\frac
{1}{2}+(2x-2y-z+t)^2}\nonumber\\
&&+\frac {1}{\sqrt 2}\sqrt {\frac
{1}{2}+(2x-2y+z-t)^2}\nonumber\\
 &&+\frac {1}{\sqrt 2}\sqrt {\frac {1}{2}+(2x+2y-z+t)^2}\nonumber\\
&&+\frac {1}{\sqrt 2}\sqrt {\frac {1}{2}+(2x+2y+z-t)^2}).
\end{eqnarray}
A little thought shows that
\begin{equation}
\sum_{\substack{i,j=1 \\ i\neq j}}^3\sum_{k,l=1}^3\sqrt {\frac
{\eta_{ik}\eta_{jl}}{
{(N-m_i)(N-m_j)}}}|\langle\chi_{ik}|\chi_{jl}\rangle| \geq\frac
{1}{3}.
\end{equation}
 The equality occurs at
\begin{equation}
   \langle\alpha|\beta\rangle=0,
   \langle\alpha|\alpha\rangle=\langle\beta|\beta\rangle=\frac{1}{2}.
\end{equation}

The average success probability $P_2$ of conclusive quantum states
sets classification \cite{WY} is
\begin{equation}\label{p2}
P_2\leq 1-\frac {1}{3}=\frac {2}{3}.
\end{equation}

From Eq.(\ref{p1}) and Eq.(\ref{p2}), we can see that Eve can not deduce $A_i$ and $B_i$ by the probability more
than $\frac{1}{3}$. Thus, no matter what kind strategy the malicious Alice $i_0$ use, she will disturb the
quantum system, make mistakes,  and therefore can be detected in M2, M3, M5, or M7. Therefore not only the
fake-signal attacking with EPR pairs but also  the fake-signal attacking with any two-particle entangled
states---general EPR pairs  can not work for the quantum secret sharing  protocol with the six states.

Note that Eve can not deduce $A_i$ and $B_i$ by the probability more than $\frac{1}{5}$ by Eq.(15) and Eq(19)
in \cite{YGL}, but in the present protocol, for getting all Alices' encoding information $A_i$ and $B_i$, Eve
must manage to distinguish nine states in Eq.(\ref{9states}) or even more. It implies that this quantum secret
sharing scheme  between multi-party and multi-party with six states is much more secure than that with four
states in \cite{YGpra1, YGL}.

Remark 4. For safety, Alice 2, Alice 3, $\cdots$, Alice $m$ must
utilize at least three of $I$, $\sigma_x$, $i\sigma_y$, and
$\sigma_z$ to encode their secret message. If they use two,  then
$\frac{1}{3}$ secret information will be leaked.

Remark 5.  Alice $i$ ($2\leq i\leq m$) applying unitary operation
$U_1$ or $U_2$ randomly on some qubits is  to achieve the aim such
that no one or part of Alice 1, $\cdots$ , Alice $m$ can extract
some information of others \cite{YGpra1}.

Remark 6. This protocol is safer than that with four states \cite{YGpra1, YGL}, which can also be shown in
section III.

 This secret sharing protocol between $m$ parties and $n$ parties is almost $100\%$ efficient as all the keys
can be used in the ideal case of no eavesdropping,  while the quantum secret sharing protocols with entanglement
states \cite {HBB} can be at most $50\%$ efficient in principle. In this protocol, quantum memory is required to
store the qubits which has been shown available in the present experiment technique \cite {GG02}. However, if no
quantum memory is employed,  all Bobs measure their qubits before  Alice $i$'s ($1 \leq i\leq m$) announcement
of basis, the efficiency of the present protocol falls to $33.33\%$.

\section {security}

 By means of the special filters, photon number splitters, single-photon detectors, the eavesdropping check
of each member Alice $i$ ($i=2,3,\cdots, m$) in group 1, inserting of decoy states in M2 and M3, unitary
operations in M2 and M3, and the random measurements of all Bobs on their respective qubits chosen at random,
either an ($m+n+1$)-th party (an "external" eavesdropper) or the dishonest agent of two groups  can be found  by
the honest agents. Therefore all Alices and all Bobs must be honest.

The encoding of secret messages  by  Alice $i$ ($1\leq i\leq m$) is identical to the process in a one-time-pad
encryption where the text is encrypted with a random key as the state of the photon in the protocol is
completely random. The great feature of a one-time-pad encryption is that as long as the key strings are truly
secret, it is  completely safe and no secret messages can be leaked even if the cipher-text is intercepted by
the eavesdropper. Here the secret sharing protocol is even more secure than the classical one-time-pad in the
sense that an eavesdropper Eve can not intercept the whole cipher-text as the photons' measuring-basis is chosen
randomly.  So the transmission of qubits between
 authorized members in the two groups is secure. Thus the security of this secret sharing protocol depends entirely
  on the second part when Alice $m$
sends the $l$-th sequence of $N$ photons to Bob $l$ ($1\leq l\leq n$).

 The process for ensuring a secure block of $nN$ qubits ($n$ secure sequences of  $N$ photons) is similar to
 that in quantum key distribution protocol based on six quantum states
 \cite{BBBW, Bru}, in the following called BBBWB  six-state protocol. The process of this secret sharing between
  $m$ parties and $n$ parties after all Alices encoding
 their respective messages using unitary operations is in fact identical to $n$ independent
 BBBWB six-state  protocol processes, which has been proven unconditional secure \cite {Lo}. Lo \cite {Lo} has
 demonstrated the unconditional security of BBBWB  six-state protocol up to a bit error rate of 12.7 percents, by
 allowing only one-way classical communications in the error correction/privacy amplification procedure between
 Alice and Bob. This shows a clear advantage of the six-state protocol over BB84 \cite{BB84}, which has been proven to be secure
 up to 11 percents, if only one-way classical communications are allowed. Lo \cite {Lo} has shown that an advantage of
  the six-state protocol lies in the Alice and Bob's ability to establish rigorously from their test
  samples
   the non-trivial mutual information between the bit-flip and phase error patterns.
 Thus the security for the present quantum  secret sharing
 between multi-party and multi-party is guaranteed.

 In summary, we propose a  scheme for quantum secret sharing between multi-party and multi-party with three conjugate
 bases, or six states, where  no
 entanglement  is employed.
In the protocol,  Alice 1 prepares a sequence of single photons in one of six different states according to her
two random classical strings, other Alice $i$ ($2\leq i\leq m$) directly encodes her two random classical
information strings on the resulting sequence of Alice $(i-1)$ via unitary operations, after that Alice $m$
sends $1/n$ of the sequence of single photons to each Bob $l$ ($1\leq l\leq n$). Each Bob $l$ measures his
photons according to all Alices' measuring-basis sequences. All Bobs must cooperate in order to infer the secret
key shared by all Alices. Neither a subset of either all Alices or all Bobs nor the union of a subset of all
Alices and a subset  of all Bobs can  extract secret information, but each entire group (each of the entire set
of all Alices and the entire set of all Bobs) can. This scheme has no secure flaws proposed in \cite{DYLLZG} and
\cite{LiChangHwang}. It is secure against the attack with invisible photons \cite{caiqingyu} and the fake signal
attack with any two-particle entangled state (the special case of which is the attack with EPR pairs
\cite{DLZZ2}),
 and safer than the one \cite{YGpra1, YGL} based
on two conjugate bases, i.e. four states, which is secure. This
shows the advantage of our proposed scheme based on six states over
Ref.\cite{YGpra1}.
      As entanglement, especially the inaccessible multi-party entangled state, is not
 necessary in the present quantum secret sharing protocol between $m$-party and $n$-party, it may be more
 applicable when the numbers $m$ and $n$ of the parties of secret sharing are large. Its theoretic efficiency is
 also doubled to approach $100\%$. This protocol is feasible with present-day technique.

 \acknowledgments This work was supported by the National Natural
Science Foundation of China under Grant No: 10671054 and Hebei Natural Science Foundation of China under  Grant
No: A2005000140 and  Natural Science Foundation of Hebei Normal University.

\end{document}